\documentclass[aps,pre,twocolumn,10pt]{revtex4-2}
\usepackage[dvipsnames]{xcolor}
\usepackage{graphicx} 
\usepackage{amsmath,amsfonts,amssymb,bm}

\usepackage{XCharter}
\usepackage[T1]{fontenc}
\usepackage{mathptmx}
\usepackage[framemethod=tikz]{mdframed}
\usepackage{amsmath,amsthm,amssymb,mathrsfs}
\usepackage{slashed}
\usepackage{graphicx}
\usepackage[export]{adjustbox}
\usepackage{float}
\usepackage{multirow, makecell}
\usepackage{hyperref}
\usepackage{enumitem}
\hypersetup{colorlinks=true,citecolor=red,linkcolor=NavyBlue,urlcolor=NavyBlue}
\usepackage[caption=false]{subfig}
 
\usepackage{natbib}

\usepackage{relsize}
\usepackage{footmisc}
\usepackage[left=1.55cm,right=1.55cm,top=1.6cm,bottom=1.56cm]{geometry}
\usepackage{tcolorbox}
\usepackage{longtable}

\newcommand{\bmat}[4]{\left(\begin{array}{cc}#1&#2\\#3&#4\end{array}\right)}

\newcommand{\av}[1]{\langle #1\rangle}

\begin{document}

\title{Origins of Instability in Dynamical Systems on Undirected Networks}

\author{Shraosi Dawn$^{*}$, Subrata Ghosh$^{\P}$, Chandrakala Meena$^{\dag}$,  Tim Rogers$^{\S}$, Chittaranjan Hens$^{*}$ } 

\affiliation { $^{*}$ Center for Computational Natural Sciences and Bioinformatics, International Institute of Information Technology Hyderabad, Gachibowli, Hyderabad-500032, Telangana, India.}
\affiliation { $^{\P}$ Division of Dynamics, Lodz University of Technology, Stefanowskiego 1/15, 90-924 Lodz, Poland}

\affiliation { $^{\dag}$ Physics Department, Indian Institute of Science Education and Research (IISER), Pune, 411008, Maharashtra, India.}

\affiliation { $^{\S}$ Department of Mathematical Sciences, University of Bath, Bath, BA27AY, UK.}

\begin{abstract}
Robustness to perturbation is a key topic in the study of complex systems occurring across a wide variety of applications from epidemiology to biochemistry. Here we analyze the eigenspectrum of the Jacobian matrices associated to a general class of networked dynamical systems, which contains information on how perturbations to a stationary state develop over time. We find that stability is always determined by a spectral outlier, but with pronounced differences to the corresponding eigenvector in different regimes. We show that, depending on model details, instability may originate in nodes of anomalously low or high degree, or may occur everywhere in the network at once. Importantly, the dependence on extremal degrees results in considerable finite-size effects with different scaling depending on the ensemble degree distribution. Our results have potentially useful applications in network monitoring to predict or prevent catastrophic failures, and we validate our analytical findings through applications to epidemic dynamics and gene regulatory systems. 
\end{abstract}

\maketitle 

\section{Introduction}
Predicting the onset of instability in large networked systems is a long-standing challenge in theoretical physics, of relevance to applications in a wide variety of fields. The linear stability of a stationary state in such systems is determined by the eigenspectrum of the {\it Jacobian} matrix, particularly by its rightmost eigenvalue. 
However, the scale and complexity of many networked dynamical systems pose a considerable challenge to determining the Jacobian. The traditional random matrix theory (RMT) paradigm treats the elements of the Jacobian as random, subject to constraints that reflect the nature of the underlying dynamical system. A large body of literature has developed adapting RMT tools to account for structural properties of Jacobians linked to different application areas, including features of ecological networks  \cite{may1972will,grilli2016modularity}, economic models  \cite{potters2005financial} and epidemic spread \cite{rogers2015assessing,moore2020predicting}. There has been some effort to look more closely at features that relate the nature of the fixed points to their stability (such as the role of abundance in ecological stability \cite{sinha2005evidence,gibbs2018effect,meena2017threshold}), and other influential work seeking to sample Jacobians from across a whole space of plausible dynamics \cite{gross2006generalized}, but the convention of exploring Jacobians in isolation has been largely unchallenged. 

In recent years, however, researchers have begun to directly investigate the behaviour of random dynamical systems, with some findings overturning previous theories. Baron and collaborators \cite{baron2022eigenvalues,baron2023breakdown} have shown how the surviving species in a random ecosystem have spectral statistics that break from those of the previously established universality class. Dynamical Mean Field Theory has been deployed to elucidate the behavior of random Lotka-Volterra models \cite{azaele2024generalized} and expanded to a more general class of models in \cite{metz2025dynamical}. 
The role of network structure and dynamics has been explored more deeply in Meena {\it et al.} \cite{meena2023emergent}, which introduces the {\it Dynamic Jacobian Ensemble} linking emergent stability in network dynamics to the statistics of node degrees and Jacobian entries. 
These results show the ways in which the stability of networked dynamical systems may be more complex than RMT models suggested. 

Here, we combine the techniques of \cite{meena2023emergent} with sparse random matrix theory \cite{metz2019spectral} to tackle the emergent stability properties of a broad class of networked dynamical systems. We study general nonlinear systems belonging to the Barzel-Barab\'{a}si family of equations:
\begin{equation}
\dfrac{dx_i}{dt}=f(x_i)+\,\sum\limits_{j=1}^{n}A_{ij}g(x_i)h(x_j)\,.
\label{eq:BBdynamics}
\end{equation}
The matrix $A$ is the adjacency matrix of the network under study, which we for now assume to be symmetric, unweighted and sparse but with a relatively high mean degree. With suitable choices of self-term ($f$) and interaction terms ($g,h$), this family of equations may represent a wide variety of phenomena from cellular to societal~\cite{gao:2012,ji:2020,kundu:2022,masuda:2022,ji:2023signal,wu2023rigorous,hatton2024diversity,ghosh2025universal,roy2024impact,saha2020infection}. 

As discovered in Ref.~\cite{meena2023emergent}, and briefly recapped in the next section, the elements of Jacobian matrices of \eqref{eq:BBdynamics} obey the general structure $J_{ii}=a_i$ and $J_{ij}=A_{ij}b_ic_j$, where $\bm{a}$, $\bm{b}$ and $\bm{c}$ are certain vectors derived from the dynamical model. This structure is amenable to analysis using the tools of sparse random matrix theory \cite{rogers2008cavity,rogers2009cavity,metz2019spectral}. Large random matrices of this form typically feature an eigenvalue bulk as well as a number of outliers. Here we focus on the case that the entries of $\bm{b}$ all have the same sign (and similarly for $\bm{c}$), so that the spectrum of $J$ is purely real \footnote{This can be shown by introducing $S_{ji}=\delta_{ij}|c_i/b_i|^{1/2}$, so that $\tilde J=SJS^{-1}$ is real and symmetric.}.
We find this case covers most applications of interest; nevertheless, this assumption could be relaxed using techniques from \cite{rogers2009cavity} for the spectral bulk. 

In what follows, we show how the cavity method \cite{mezard1987spin} can be applied to elucidate the origins of instability in sparse networked dynamical systems of the form \eqref{eq:BBdynamics}. We find that the rightmost eigenvalue is always outside of the spectral bulk, an important point of contrast with May's famous result \cite{may1972will} and many related works~\cite{sommers1988spectrum,pettersson2020predicting,tarnowski2020universal,grela2017drives}. 
Unlike prior work, we are able to distinctly categorize the stability-determining outliers into two classes; those determined by the average value of both the diagonal and off-diagonal elements of the {\it Jacobian} matrix, and those stemming from the minimum or maximum degree nodes. Depending on the parameters of the dynamical system under study, the associated leading eigenvector might be localized on particular nodes, or delocalized such that large-scale fluctuations may occur across the whole network. We demonstrate the generality and practical relevance of our analytical framework by applying it to two nonlinear systems of broad interest, namely the SIS epidemic model and gene regulatory dynamics.

\section{Cavity Method for Dynamical Spectral Outliers}

\subsection{Jacobian matrices of Barzel-Barab\'asi dynamics}
\label{Jac_constraction}
Consider the generic networked dynamical system as in \eqref{eq:BBdynamics}

where $A$ is the adjacency matrix of a large symmetric network. We are interested in the case that the average degree of nodes in the network is large, so that the sum above is approximately self-averaging. We make a heterogeneous mean-field assumption neglecting degree-degree correlations to write
\begin{equation}
    \dfrac{dx_i}{dt}\approx f(x_i)+ g(x^*_i)d_i \av{h}\,,
\end{equation}
where $d_i$ is the degree of node $i$ and $\av{\cdots}$ denotes the average over the network. Steady states of these dynamics are determined by solutions to the equation 
\begin{equation}
    f(x^*_i) + g(x^*_i) d_i \langle h \rangle = 0.
\end{equation}
The general structure of the {\it Jacobian} matrix around a fixed point $\bm{x}^\star$ is then as follows:
\begin{align}
    J_{ii} &= f'(x^*_i) + g'(x^*_i) d_i \langle h \rangle, \\
    J_{ij} &= A_{ij} g(x^*_i) h'(x^*_j).
\end{align}

\subsection{Spectral density and the resolvent}
\label{spec_density}

Motivated by the above application to complex dynamical systems, we study the spectrum of matrices of the form 
\begin{equation} 
\begin{split}
J_{ii} = a_{i}, \quad J_{ij} = A_{ij}b_{i}c_{j}\,.
\end{split}
\label{jacobian_gen}
\end{equation}
Here $A$ is the adjacency matrix as above and $\bm{a}$, $\bm{b}$, $\bm{c}$ are vectors of parameters. We assume that all entries of $\bm{b}$ are non-zero and have the same sign, and similarly for $\bm{c}$. 

We can write $\{\lambda_i\}_{i=1}^N$ for the eigenvalues of $J$. Introduce the invertible diagonal matrix $S_{ij}=\delta_{ij}|c_i/b_i|^{1/2}$ and $\tilde J=SJS^{-1}$. The transformed matrix $\tilde J$ has entries
\begin{equation}
    \tilde J_{ii} = a_{i}|c_i/b_i|, \quad \tilde J_{ij} = \text{sign}(b_ic_j)A_{ij}\sqrt{|b_{i}b_{j}c_ic_j|}=\tilde{J}_{ji}\,.
\end{equation}
Since $\tilde J$ is real and symmetric, we conclude that $J$ has a purely real spectrum. 

Write $\rho(\lambda)=N^{-1}\sum_i\delta(\lambda-\lambda_i)$ for the spectral density of $J$. The resolvent matrix  $G = (J - \lambda-i\eta)^{-1}$ recovers the spectral density of $J$ in the usual  way via
\begin{equation}
    \rho(\lambda) = \lim_{\eta\searrow0}\frac{1}{N\pi}\text{Im}\,\text{Tr}\, G\,.
    \label{rholambda}
\end{equation}
In the limit of large networks ($N\to\infty$) it is expected that the spectrum of $J$ will be composed of a continuous part known as the bulk, plus a number of outlier eigenvalues.

In what follows we will apply the cavity method \cite{rogers2008cavity,neri2016eigenvalue,metz2019spectral} to compute approximations to $G$ in different regimes. We begin by expressing the elements of $G$ in an expansion around a particular node $i$, which for convenience we locate as the first row/column. Using standard formulae for block matrix inversion, we have
\begin{equation}
      G=\bmat{G_i}{-G_iJ_{i\ast}G^{(i)}}{-G_iG^{(i)}J_{\ast i}}{G^{(i)}+G^{(i)}J_{\ast i}G_iJ_{i\ast}G^{(i)}}\,,
      \label{Gpart}
\end{equation}
where we use the superscript $(i)$ to denote quantities computed for the `cavity' network with node $i$ removed. The top left element here obeys
\begin{equation}
        G_{i}=\left(a_{i}-\lambda-i\eta-\sum_{j}b_{i}b_{j}c_{i}c_{j}G_{j}^{(i)}\right)^{-1}\,.
    \label{green_fun}
\end{equation}

\subsection{Bulk} 
\label{appendix:bulk}

The bulk is identified as the range of $\lambda$ for which the right hand side of \eqref{rholambda} is non-vanishing. Assuming as above that the mean degree is large and the network mostly decorrelated, we will pursue a mean field approximation replacing all terms in \eqref{green_fun} with their average. Specifically,
\begin{equation}
  \av{G}  =  \Big(\av{a} - \lambda - \av{b}^2 \av{c}^2 \av{d} \av{G}\Big)^{-1}.
  \label{G_inv_1}
\end{equation}
This is a quadratic equation for the spectral bulk $\av{G}$:
\begin{equation*}
       \av{b}^2 \av{c}^2 \av{d} \av{G}^2 - (\av{a} - \lambda)\av{G} + 1 =0.
\end{equation*}
The bulk edge is found when the discriminant vanishes,
\begin{equation*}
    (\av{a} - \lambda)^2 - 4\av{b}^2 \av{c}^2 \av{d} =0,
\end{equation*}
\begin{equation}
    \Rightarrow\quad \lambda_{\text{bulk}} = \av{a} + 2\av{b}^2 \av{c}^2 \av{d}^{1/2}.
\end{equation}
Note that we only consider the positive root as we are concerned with the  stability of the system \eqref{eq:BBdynamics} and thus in the rightmost eigenvalues of $J$.
\subsubsection{ Outliers} 
\label{appendix:outlier}

Outlier eigenvalues are encoded in $G$ by examination of the relation 
\begin{equation}
    G=\sum_i\frac{\bm{v}_i\bm{v}_i^T}{\lambda_i-\lambda-i\eta}\,, 
\end{equation}
where $\bm{v}_i$ is the unit-length eigenvector associated to eigenvalue $\lambda_i$. If $\lambda$ is an outlier eigenvalue with corresponding eigenvector $\bm{v}$, then the sum is dominated by a single term and 
\begin{equation}
    G=\eta^{-1}\bm{v}\bm{v}^T+\mathcal{O}(1)
    \label{Geta}
\end{equation}
Summing over the entries of the first row of $G$ we have two alternative expressions. From \eqref{Geta} we have
\begin{equation}
    \sum_jG_{ij}\approx\frac{\sigma}{\eta}v_i\,,
    \label{Gsum1}
\end{equation}
where $\sigma=\sum_jv_j$. Summing the top row of \eqref{Gpart}, however, gives
\begin{equation}
    \sum_jG_{ij}=G_i-G_i\sum_{j,k}J_{ij}G^{(i)}_{jk}=G_i\left(1-\sum_{j,k}A_{ij}b_ic_jG^{(i)}_{j,k}\right)\,.
    \label{Gsum2}
\end{equation}
Introducing $v_j^{(i)}=\eta\sum_kG_{jk}^{(i)}$ and $\sigma^{(i)}=\sum_jv_j^{(i)}$ then equating \eqref{Gsum1} and \eqref{Gsum2} yields 
\begin{equation}
    \frac{\sigma}{\eta} v_i = G_i \left(1-\frac{\sigma^{(i)}}{\eta}\sum_jA_{ij}b_{i}c_{j}v^{(i)}_j\right)\,.
    \label{Gout}
\end{equation}
From here there are two possibilities of interest, depending on whether $\bm{v}$ has mass spread throughout the whole network (corresponding to $G_i$ being order one \cite{thouless1972relation}), or is localized near node $i$ (when $G_i$ is order $\eta^{-1})$.

{\it Delocalized outliers.}
If $\bm{v}$ is delocalized then $G_i$ is order one and $\sigma^{(i)}\approx \sigma$. The equation for the leading order term in \eqref{Gout} is then
\begin{equation}
    v_i = -G_i \sum_jA_{ij}b_{i}c_{j}v^{(i)}_j\,.
     \label{cavity_v}
\end{equation}
Making the same mean-field approximation as for the bulk we arrive at 
\begin{equation}
    \av{v} =-  \av{G} \av{b} \av{c} \av{d} \av{v} ,
    \label{eig_vec}
\end{equation}
hence
\begin{equation*}
     \av{G} = - \frac{1}{\av{b} \av{c} \av{d}}.   
\end{equation*} 
When $\av{v}$ is nonzero, we can determine the $\av{G}$  by solving the relevant equation. Substituting this solution into (\ref{G_inv_1}) allows us to derive an expression for large $\av{d}$:
\begin{equation}
    \lambda_o = \av{a} + \av{b} \av{c} (\av{d} +1)\approx \av{a} + \av{b} \av{c} \av{d}.
    \label{z_o}
\end{equation} 

Note that this value is always at least as big as $\lambda_{\text{bulk}}$, so for the systems under study here we expect stability to always be determined by a spectral outlier, in contrast to what is observed in much of the existing literature focused on random Jacobians without structure imposed by an underlying dynamical system. 

The corresponding (un-normalized) eigenvector (found using \ref{cavity_v}) in this case has entries
\begin{equation}
    v_i=\frac{d_ib_i\av{c}}{\av{a}+\av{b}\av{c}\av{d}-a_{i}+d_i\av{b}^2\av{c}^2\av{G}_{\text{bulk}}}\,.
\end{equation}

{\it Localized outliers.}
For spectral outliers localized on nodes we have that $G_i$ is of order $1/\eta$ and hence to leading order \eqref{Gout} becomes simply
\begin{equation}
    \frac{\sigma}{\eta} v_i = G_i \,.
\end{equation}
For $G_i$ to be order $1/\eta$, from \eqref{green_fun} we require that 
\begin{equation}
    \lambda = \lambda_i\approx a_i - \sum_jA_{ij}b_ib_jc_ic_jG_j^{(i)}\,.
\end{equation} 
We make the approximation that $G_j^{(i)}\approx\av{G}$ so that in the mean-field we must simultaneously solve:
\begin{equation}
    \lambda_i= a_i - d_i\av{b}^2\av{c}^2\av{G}
    \label{z_i_gen_7}
\end{equation}
and
\begin{equation*}
     \av{G}^{-1}  =  \av{a} - a_i + d_i \av{b}^2 \av{c}^2 \av{G} - \av{b}^2 \av{c}^2 \av{d} \av{G},
\end{equation*}
\begin{equation*}
      \av{b}^2 \av{c}^2 (d_i  -  \av{d}) \av{G}^2 + (\av{a} - a_i)\av{G} -1 = 0, 
\end{equation*}
\begin{equation}
    \av{G} = \frac{- (\av{a} - a_i) \pm \sqrt{(\av{a} - a_i)^2 + 4\av{b}^2 \av{c}^2 (d_i  -  \av{d})} }{2\av{b}^2 \av{c}^2 (d_i  -  \av{d})}.
    \label{G_avh_8}
\end{equation}
Replacing the above equation in eqn. (\ref{z_i_gen_7}) and substitute $\lambda$ with $\lambda_i$ we have:
\begin{equation*}
    \lambda_i = a_i - d_i \av{b}^2 \av{c}^2 \bigg[\frac{- (\av{a} - a_i) \pm \sqrt{(\av{a} - a_i)^2 + 4\av{b}^2 \av{c}^2 (d_i  -  \av{d})} }{2\av{b}^2 \av{c}^2 (d_i  -  \av{d})}\bigg],
\end{equation*}
\begin{equation}
    \lambda_i = a_i - d_i\frac{(\av{a} - a_i)}{2(d_i  -  \av{d})} \pm d_i \bigg[\frac{ \sqrt{(\av{a} - a_i)^2 + 4\av{b}^2 \av{c}^2 (d_i  -  \av{d})} }{2 (d_i  -  \av{d})}\bigg].
    \label{z_i_deloca}
\end{equation}

\section{Applications}
\subsection{Dynamic Jacobian Ensemble}
\begin{figure}[h]
    \centering
    \includegraphics[width=\linewidth]{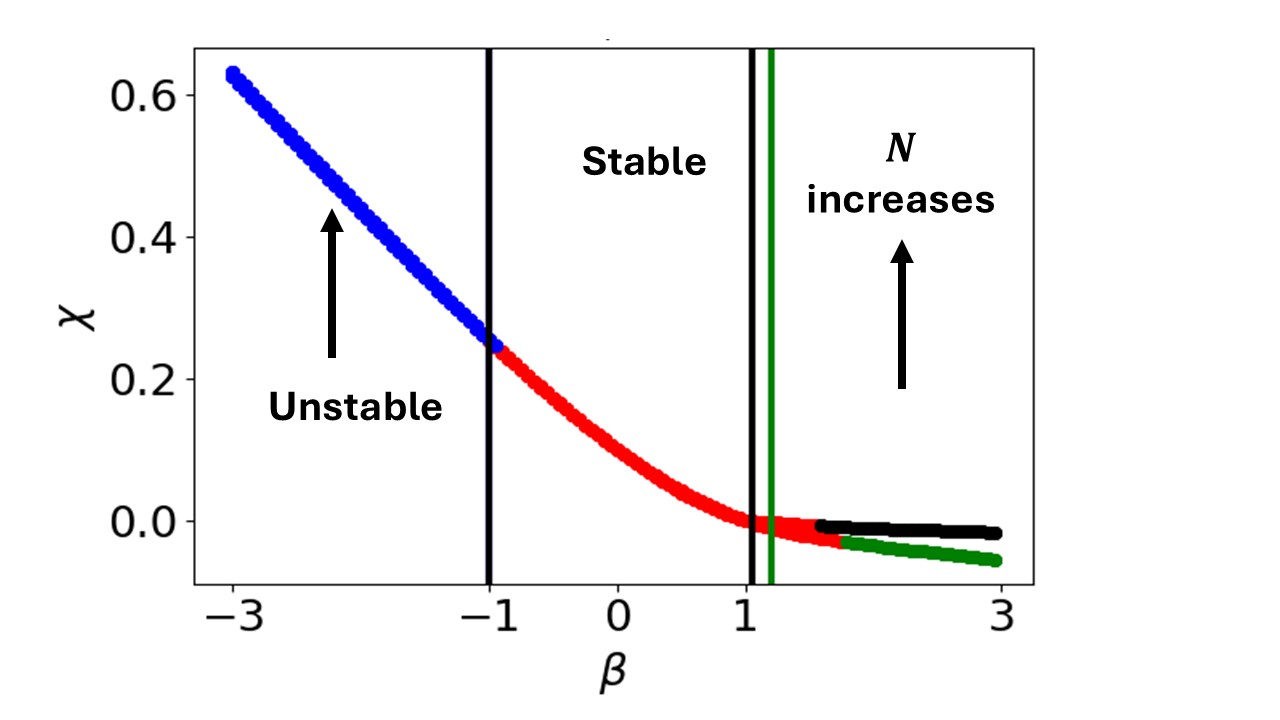}
    \caption{Stability of the dynamic Jacobian ensembles on a Barab{\'a}si-Albert (BA) network with two different network sizes: $N=200$, and $N=1500$, with $\langle d \rangle=10$ and with a fixed minimum degree of $5$. The maximum degree is $\approx 53$ for $N=200$ and $179$ for $N = 1500$. 
   In BA networks, increasing the network size results in a higher maximum degree, which consequently reduces the extent of the stable regime, as indicated by the bounded region between the vertical black lines. The numerical and theoretical $\lambda_{\rm max}$  have been plotted in the {\color{blue} (Appendix \ref{additional figures}, and, Fig.\ \ref{fig:enter-label})}, for three $\beta$  values: $-2, 0, 2$. }     
    \label{fig:dynamic_jacobian}
\end{figure}

Meena {\it et al.} discovered a degree-based characterization of nonlinear dynamical systems connecting over complex networks~\cite{meena2023emergent}. The resultant Jacobian ensemble is characterized by dynamic exponents $(\mu,\nu,\rho)$ that are derived from the behavior of the functions $f,g,h$ near the stationary state $x^\star$ and degrees $d$ of network nodes. Specifically, we examine the Jacobian matrix 
of the following form
\begin{equation}
    \begin{split}
    J_{ii}&=-\beta-\chi d_i^{\mu}\\
    J_{ij}&=A_{ij}d_i^{\nu}d_j^{\rho}\,,
    \end{split}
\end{equation}
where $\beta$ and $\chi$ control the intrinsic and network-driven self-interaction terms. Using equations   \eqref{z_o}
\eqref{z_i_deloca} 
we find expressions for outlier eigenvalues, to leading order:
\begin{equation}
\begin{split}
    \lambda_{\circ}&=-\beta-\chi\av{d}^\mu+\av{d}^{1+\nu+\rho}\\
    \lambda_{i}&=-\beta-\chi d_{i}^\mu\,.\\
    \label{lambda_meena}
\end{split}
\end{equation}
The resulting phase diagram has six parts, depending on which eigenvalue is largest, and whether it is positive or not. The stability boundary is determined to be
\begin{equation}
    \beta_c=\begin{cases}
        -\chi \max\{d_i^\mu\}&\chi<\frac{\av{d}^{1+\nu+\rho}}{\av{d}^\mu-\max\{d_i^\mu\}}\\
        -\chi \min\{d_i^\mu\}&\chi>\frac{\av{d}^{1+\nu+\rho}}{\av{d}^\mu-\min\{d_i^\mu\}}\\
        -\chi\av{d}^\mu+\av{d}^{1+\nu+\rho}&\text{otherwise.}
    \end{cases}
    \label{beta_c}
\end{equation}
In Figure \ref{fig:dynamic_jacobian},  we illustrate the stability boundary (i.e. the rightmost eigenvalue) on a Barab{\'a}si-Albert network \cite{barabasi1999emergence} with dynamical exponents $\mu=1, \nu=-1, \rho=0$. 
For a range of negative $\beta$, the least stable eigenvalue is localized on the minimum degree nodes (blue curve); in the parameter region above the blue line the system is stable. The red curves correspond to the points where the delocalized analytical formula crosses the zero eigenvalue, which determines stability for intermediate values of $\beta$. For larger positive $\beta$ stability is determined by the maximum degree node, and the boundary is marked in black (networks size $N=1500$) and green ($N=200$). The vertical lines indicate the transition points where the prediction of the largest eigenvalue switches between localized and delocalized regimes.

\begin{figure}
    \centering
    \includegraphics[width=1.0\linewidth]{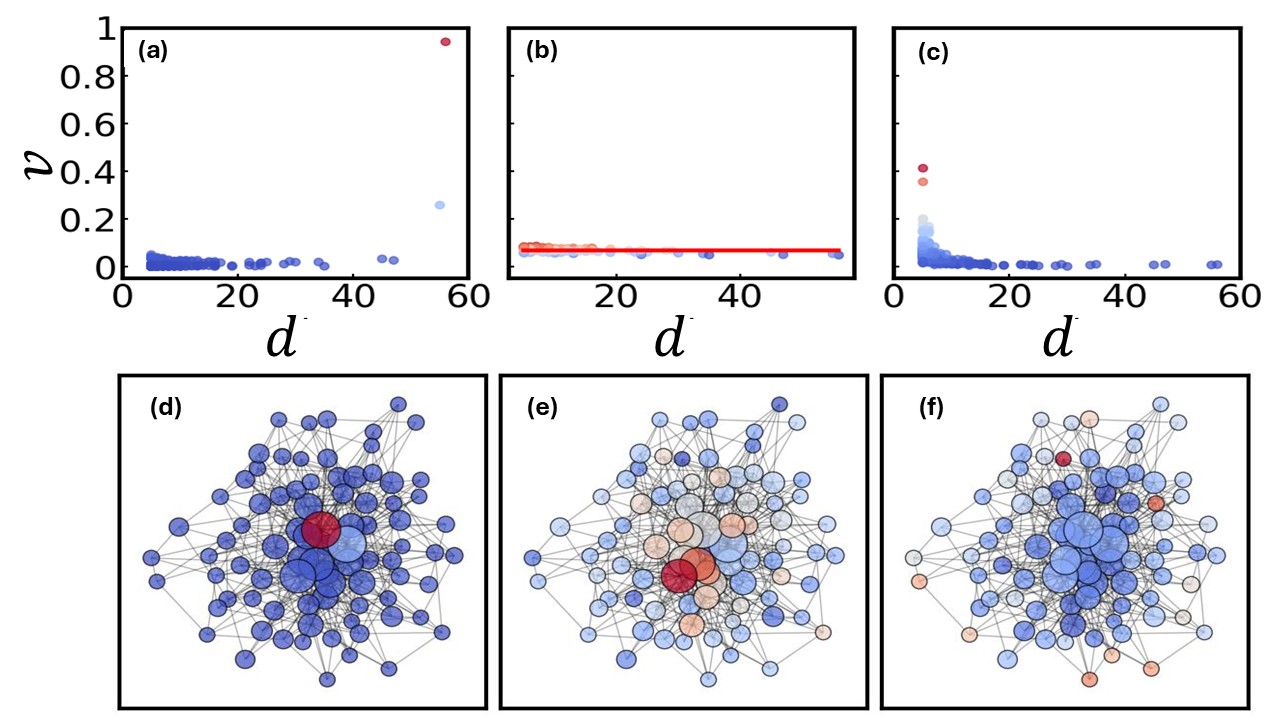}
    \caption{Principal eigenvector of the dynamic Jacobian ensemble on a BA network of size $N=200$ (same BA graph as in Figure 1), with dynamical exponents $\mu=1,\nu=-1$ and $\rho=0$, for the fixed $\beta=0$ and three values of $\chi$ = -0.1, 0.01, 0.3 (left to right). 
    The size of the circles is scaled with their degree, and the color is scaled with the value of the elements of the leading eigenvector, from high (red) to low (blue).    }
    \label{fig:evcvalue}
\end{figure}

In the top row of Fig.\  \ref{fig:evcvalue} we report the eigenvector entries as a function of  network degree, while the second row illustrates the network structure. Reading across the panels we observe as $\beta$ increases the shift in the least-stable eigenvectors being localized on nodes of high degree, to delocalized, to localized again on low degree nodes.


\subsection{Epidemic dynamics}
The Susceptible-Infectious-Susceptible (SIS) model of disease dynamics \cite{krause2018stochastic,hens2019spatiotemporal} considers network nodes as populations of individuals, for example in different residential zones, and tracks changes in the proportion $x_i$ of infectious individuals at each node. This model is realized in the Barzel-Barab\'{a}si framework by setting $f(x)=-x$, $g(x)=\beta(1-x)$ and $h(x)=x$, where $\beta>0$ 
is the rate of infection. Assuming large mean degree, the disease-extinct fixed point $x_i\equiv0$ has rightmost eigenvalue $\lambda_{\text{ext}}=\beta\av{d}-1$. If $\beta>1/\av{d}$ this fixed point destabilizes, and an outbreak is possible. For the disease-endemic fixed point, we compute the approximate expressions  
\begin{equation}
    x^\star_{i}=1-\frac{1}{1+d_i\beta\av{x}}\,, \quad \av{x^\star}=1-\frac{1}{\av{d}\beta}\,.
\end{equation}
Again, following equations  \eqref{z_i_deloca} and \eqref{z_o}, one can write the outlier eigenvalues to leading order in large mean degree (see  {\color{blue} Appendix \ref{SIS}}  for detailed calculation) as

\begin{equation}
\begin{split}
    \lambda_{\circ}&=1-\beta\av{d}\\
    \lambda_{i}&\approx d_i/\av{d}-1-\beta d_i\,.\\
\end{split}
\label{SIS_lambda}
\end{equation}

\begin{figure}
    \centering
    \includegraphics[width=0.8\linewidth]{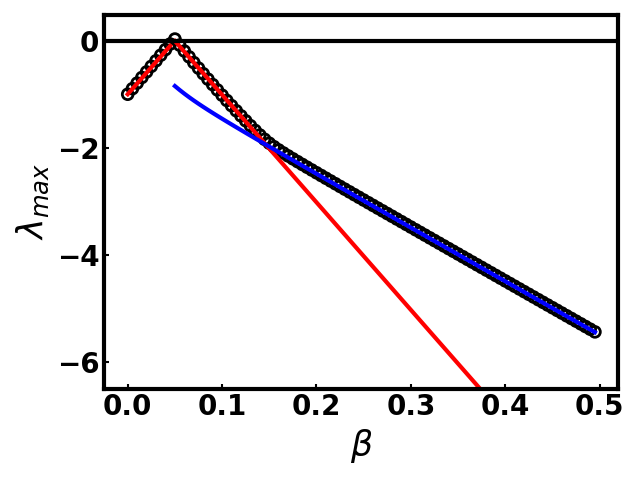}
    \caption{The largest eigenvalue is plotted as a function of the infection parameter $\beta$ for the random network of SIS dynamics. The maximum degree of the network is $35$, the minimum degree is $10$, and the average degree ($\langle d \rangle$) is fixed at $20$. The black circles are obtained from the numerical data.  The zero fixed point is stable until $0.05$, and the associated theoretical line is predicted by $\lambda_{\text{ext}}=\beta\av{d}-1$, and marked by the red line. Beyond $\beta=0.05$, the largest localized (blue line) and delocalized (red line)  eigenvalues were predicted by the Eqn.\ (\ref{SIS_lambda}).  }
    \label{fig:sis_plotn}
\end{figure}

When the infection rate $\beta$ is varied, we find a cross-over  between localization and delocalization of the rightmost eigenvector, illustrated in Figure ~\ref{fig:sis_plotn}. For $\beta<0.15$, the $\lambda_{max}$  is determined by $\lambda_0$ of eqn. \eqref{SIS_lambda}, whereas for larger values it is determined by minimum degree of the network. In {\color{blue} Appendix \ref{additional figures} (Fig.\  \ref{fig:SIS_time_sig_inf})}, we report results on the average infection  ($\langle x^*\rangle$) as a function of $\beta$ for four synthetic networks.

\subsection{Regulatory network}
\begin{figure}
    \centering
    \includegraphics[width=0.8\linewidth]{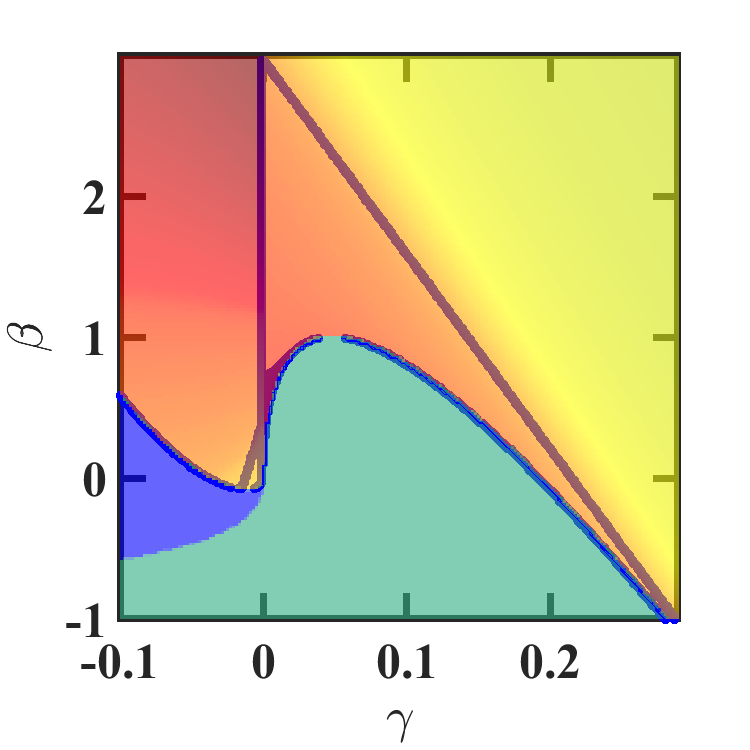}
    \caption{Eigenvalue boundary of the regulatory dynamics. The fixed points become imaginary at the green regime.  In the blue regime, the fixed points are unstable as $\lambda_{\rm max}$ is positive. The lines are marked according to the theoretically predicted  Eqns.\ (\ref{lambda_zero_Regu}-\ref{lambda_i_Regu}).  In the blue and  deep green regime, the system is unstable. The other colors (yellow to red)  are marked according to the weighted sum of the degree vector and  the  leading eigenvector. In the deep red regime, the large degree ($d_{\rm max}$) plays the key role in determining the largest eigenvalue. In the yellow regime, the smallest degree ($d_{\rm min}$) determines the $\lambda_{\max}$. In the middle, the average degree ($\langle d \rangle $) plays the key role.}
    \label{fig:regu_phasespace}
\end{figure}

For a further example, we consider a Michaelis-Menten type dynamic extensively utilized in the modelling of genetic regulation networks \cite{karlebach2008modelling,gao2016universal, meena2023emergent,hens2019spatiotemporal,bao2022impact,ji2023signal,kauffman2004ensemble}. The relevant functional forms are $f(x)=\beta x - x^2$, $g(x)=\gamma x$, $h(x)=1-1/x$. Here $\beta$ is a self-excitation term, and $\gamma$ modulates the role of the network. Solving for the fixed points with large mean degree, we find
\begin{equation}
\begin{split}
    \av{x^\star}&=\frac{1}{2}\left(\beta + \gamma \av{d} +\sqrt{(\beta+\gamma\av{d})^2-4\gamma\av{d}}\right) \,,\\
    x^\star_i&= \beta+\gamma\left(1-\frac{1}{\av{x}}\right) d_i\,.
\end{split}   
\end{equation}
From this we identify the vector entries $a_i=-x^\star_i$, $b_i=\gamma x^\star_i$ and $c_j=(x^\star_j)^{-2}$. The delocalized outlier is computed to be 
\begin{equation}
    \lambda_\circ=-\av{x^\star}+\gamma\av{d}/\av{x^\star}
    \label{lambda_zero_Regu}
\end{equation}
while the localized eigenvalues take the form
\begin{equation}
\begin{split}
    \lambda_i = -x^\star_i &+ \frac{ d_i (\langle x^\star \rangle - x^\star_i )}{2(d_i - \langle d \rangle)} \\&\pm \frac{d_i \sqrt{(\langle x^\star \rangle - x^\star_i )^2 - 4\gamma^2 \av{x^\star}^{-2}(\langle d \rangle - d_i)}}{2(d_i - \langle d \rangle)}\,.
    \end{split}
      \label{lambda_i_Regu}
\end{equation}
As shown in Figure \ref{fig:regu_phasespace} the phase diagram of this system is complex, featuring regions with localized and delocalized stability boundaries. The deep blue and green regime delineates the unstable regime. The lines separate the contributions of $d_{\rm max}, \langle d \rangle$, and $d_{\min}$. The yellow part is determined solely by  $d_{\min}$. In the yellow part, if the system is disturbed/perturbed the states in the minimum degree node will take more time  to return  to its original position where as, in the deep red regime, the low degree nodes are more resilient, as the system's largest eigenvalue is determined by the maximum degree.  We have further investigated three more graphs (one ER, and the other two are generated from BA algorithm with two different $\langle d \rangle$). In Appendix (see {\color{blue}  Appendix \ref{regu_net}, and Appendix \ref{additional figures}}, we have investigated the regulatory networks on these graphs and checked how feasible equilibrium points evolve as system parameters (Fig.\ \ref{fig:Regu_phasespace}).
The theoretical and numerical $\lambda_{\rm max}$ are also reported  (Fig.\ \ref{fig:fraction_node})in the appendix to check the consistency.

\section{Conclusion}
We have explored the intricate dynamics that underpin the stability of complex networks. By integrating the Barzel-Barabási family of dynamical systems with approaches from sparse random matrix theory, we have shown how emergent stability in networked dynamical systems is fundamentally linked to the local statistical properties of nodes. The Dynamic Jacobian Ensemble \cite{meena2023emergent} captures the qualitative features of possible regimes, with overall network stability determined by outlier eigenvalues that may be delocalized, or may be localized on nodes of anomalously high or low degree. The application of our theoretical framework to various nonlinear systems, ranging from disease propagation in social networks  to gene regulation at the subcellular level, demonstrates its broad applicability and potential for uncovering universal principles. 

Future research should aim to relax some of the assumptions made in this study, such as the sign consistency of the entries in the vectors $\bm{b}$ and $\bm{c}$, to explore a wider range of dynamical behaviours. Additionally, empirical validation of our theoretical predictions through experimental data will be essential to solidify the practical relevance of our findings. Overall, our work contributes to a deeper understanding of the mechanisms that sustain resilient network dynamics, offering valuable insights for the design, monitoring, and protection of complex systems. 

\section{Acknowledgments}
TR thanks Joseph Baron for useful discussions. 
CM acknowledges support from ARNF India (Grant Number SRG/2023/001846). CH
acknowledges support from ARNF India (Grant Number
ANRF/ECRG/2024/000207/PMS). 

\appendix

\section{\textbf{SIS Dynamics}}
\label{SIS}
We start with the general dynamics: 
\begin{equation}
    \frac{dx_i}{dt} = f(x_i) + \sum_{j} A_{ij} g(x_i) h(x_j).
    \label{gen_dy_eqn}
\end{equation}
The values of $f(x_i)$, $g(x_i)$ and $h(x_j)$ for SIS dynamics are : $-x_i$, $\beta(1-x_i)$, and $x_j$.
The SIS dynamics possess a trivial fixed point $x^* = (0,...,0)^T$, representing zero infection, along with a non-zero fixed point. The nontrivial steady states are determined by the equation 
\begin{equation}
    f(x^*_i) + g(x^*_i) d_i \langle h \rangle = 0.
\end{equation}

Thus the corresponding nontrivial fixed points of the dynamics are 

\begin{equation*}
    - \langle x ^\star\rangle + \beta(1- \langle x ^\star \rangle)\langle x ^\star\rangle \av{d} =0 
\end{equation*}
\begin{equation}
    \langle x ^\star\rangle = 1 - \frac{1}{\beta \av{d}}
    \label{x_avg}
\end{equation}
\begin{equation*}
     - x_i ^\star  + \beta (1- x_i ^\star)\langle x^\star \rangle d_i = 0.
\end{equation*}
Thus, \begin{equation}
    x_i^* = \frac{\beta \langle x \rangle d_i }{1 + \beta \langle x \rangle d_i }.
    \label{x_i_fx_pt}
\end{equation}

The general structure of the {\it Jacobian} matrix elements corresponding to the dynamical equation given in Eq.\ (\ref{gen_dy_eqn}) is as follows:
\begin{align}
    J_{ii} &= f'(x^*_i) + g'(x^*_i) d_i \langle h \rangle, \\
    J_{ij} &= A_{ij} g(x^*_i) h'(x^*_j).
\end{align}

The relevant {\it Jacobian} matrix  for the non-zero fixed point of the SIS dynamics is 
\begin{align}
    J_{ii} = -1 - \beta \langle x^\star \rangle d_i,\\ 
    J_{ij} = \beta (1- x^\star_i)A_{ij}  ,  
\end{align}
To assess the stability of a disease-free (extinct) state, we examine the stability of the trivial fixed point. The theoretical formulation that determines the largest eigenvalue associated with this fixed point is expressed as follows:
\begin{equation}
    \lambda_{\text{extinct}} = -1 + \beta \av{d}.
\end{equation}

\par {\it Analytical formula to calculate the largest eigenvalue for the nonzero fixed point.
}\\
From the Eqn.\ \eqref{z_o}, in main text, one can write leading eigenvalue for the SIS model as:
\begin{equation*}
    \lambda_o = -1 - \beta (1 - \frac{1}{\beta \av{d}}) \av{d} + \beta \frac{1}{\beta \av{d}} \av{d},
\end{equation*}
\begin{equation}
    \lambda_o = 1 - \beta  \av{d}.
    \label{sis_z0}
\end{equation}
The localized eigenvalues (Eqn.\ \eqref{z_i_deloca}) can be written as
\begin{align*}
    \lambda_i = -1 - \langle x \rangle d_i - \frac{\langle x \rangle(-  \av{d} + d_i)}{2 (d_i - \av{d})} \pm \nonumber \\ \frac{d_i \sqrt{\beta^2 \langle x \rangle^2(-  \av{d} + d_i)^2 - 4 \beta^2 (1 - \langle x \rangle)^2(d_i - \av{d})}}{ 2(d_i - \av{d}) }.
\end{align*}
\begin{figure*}
    \centering
\includegraphics[width=16cm,height=9cm]{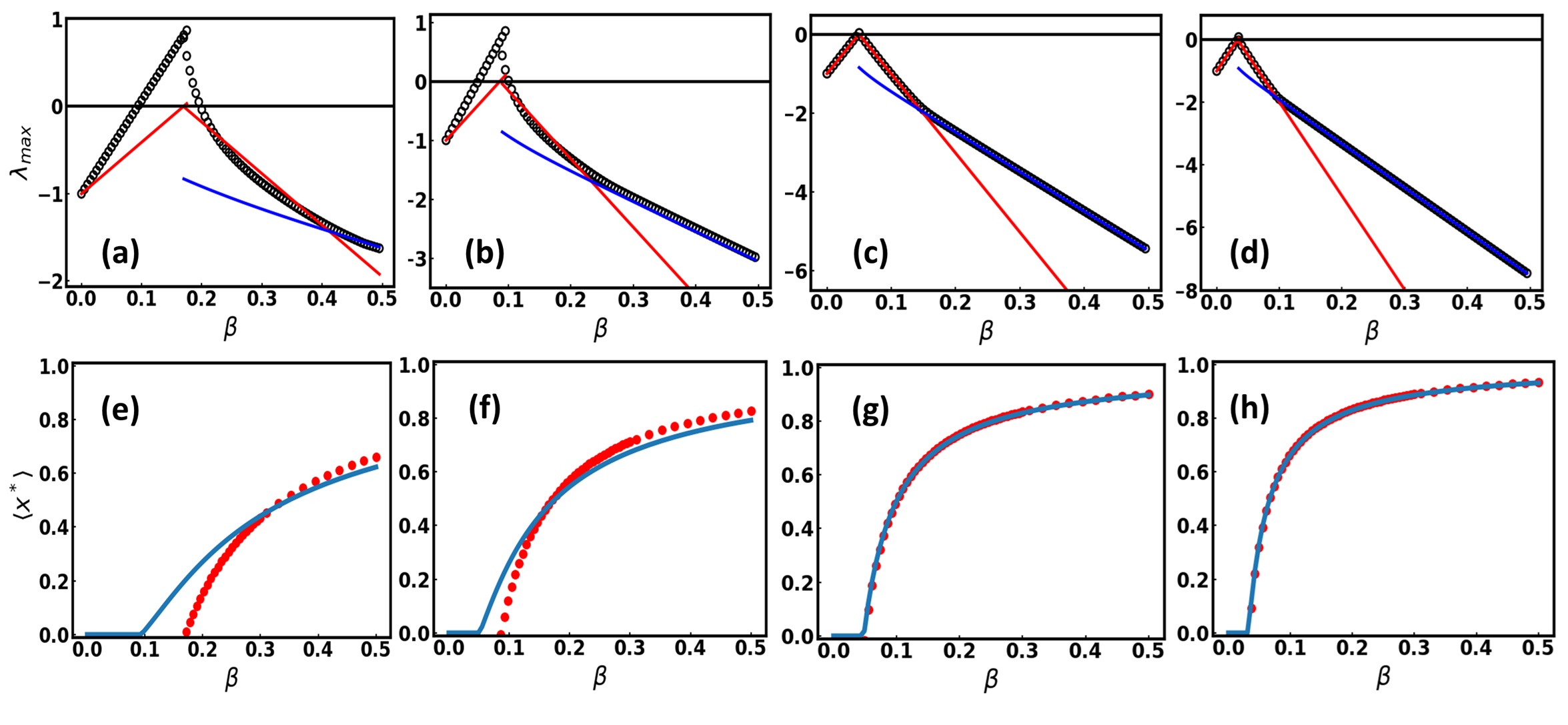}
    \caption{ {\bf SIS dynamics}. (a)-(d) Stability  ($\lambda_{\rm max}$) is plotted with a  gradual increament of \ $\beta$. (e)-(h) The mean infection  ($\langle x^* \rangle$) is reported as a function of  $\beta$. 
    $\langle x^* \rangle$ is obtained numerically (red dots in (e-h)) from the SIS model using the Runge-Kutta 4 algorithm.
    (a),(e) Network BA1   ($\av{d}=6$) is used here. 
         Here the critical onset of infection does not map  with the theoretically obtained value (blue lines), but in higher infection, they almost match each other. 
  (b), (f) BA network with average $\av{d} =12$. The theoretical and analytical results are closely  matched here.
In    (c), (g), and (d), (h) networks ER1 and ER2 are used here. (a)-(d) The black circles represent results from numerical simulations, while the red and blue lines correspond to analytical estimates using the average and minimum degrees of the network, respectively, based on \(\lambda_o\) and \(\lambda_i\) in Eqn. \eqref{SIS_lambda} in the main text. Also see Eqn.\ (\ref{sis_z0}), and (\ref{sis_zi}) of Appendix.\ \ref{SIS}.    }   \label{fig:SIS_time_sig_inf}
\end{figure*}
Further simplification leads to
\begin{align}
    \lambda_i = -1 - (1 - \frac{1}{\beta \av{d}}) d_i - \frac{\langle x \rangle}{2} \pm \nonumber \\\frac{d_i \sqrt{\beta^2 (1 - \frac{1}{\beta \av{d}})^2(-  \av{d} + d_i)^2 - 4  (\frac{1}{ \av{d}})^2(d_i - \av{d})}}{ 2(d_i - \av{d}) } .
    \label{sis_zi}
\end{align}
In the main text we report only the leading order result in the limit $\av{d}\gg1$ with $d_i/\av{d}$ assumed to be order one. Expansion of \eqref{sis_zi} in large $\av{d}$ gives
\begin{equation}
    \lambda_i=\frac{d_i}{\av{d}}-1-\beta d_i+\mathcal{O}(\av{d}^{-2})\,.
\end{equation}

\section{\textbf{Regulatory network}}
\label{regu_net}
The regulatory dynamics possess a trivial fixed point $x^* = (0,...,0)^T$, representing cell death, along with a non-zero fixed point. The nontrivial steady states are determined by the equation 
\begin{equation}
    f(x^*_i) + g(x^*_i) d_i \langle h \rangle = 0.
\end{equation}
The values of $f(x_i)$, $g(x_i)$ and $h(x_j)$ for Regulatory dynamics are :
\begin{align*}
    f(x) = \beta x - x^2, \quad
    g(x) = \gamma x, \quad
    h(x) = 1 - \frac{1}{x}.
\end{align*}
Nevertheless, the non-zero steady state can be determined and is expressed as follows:
\begin{equation}
    \langle x ^\star \rangle = \frac{\beta + \gamma \av{d} + \sqrt{(\beta + \gamma \av{d})^2 - 4\gamma \av{d}}}{2}.
\end{equation}
\begin{equation}
    x_i^\star = \beta + \gamma \left(1 - \frac{1}{\langle x \rangle} \right) d_i.
\end{equation}

Computing the derivatives with respect to time for Regulatory dynamics, we have:
\begin{align*}
    f'(x) = \beta - 2x, \quad
    g'(x) = \gamma, \quad
    h'(x) = \frac{1}{x^2}.
\end{align*}

Accordingly, the expressions obtained for the Jacobian elements in the context of regulatory dynamics are given below:
\begin{align}
    J_{ii} &= - x^*_i \\
    J_{ij} &= A_{ij} \gamma x^*_i \frac{1}{(x^*_j)^2}.
\end{align}
From the main text we have $a=- x^*_i $, $b= \gamma x^*_i$ and $c=  (x^*_j)^{-2}$. 
The theoretical expressions for $\lambda_o$ and $\lambda_i$, representing the spatial positions of the localized and delocalized outliers, are obtained by substituting the above values into equations  \eqref{z_i_deloca} and \eqref{z_o}, in main text:
\begin{equation}
      \label{Regu_zo}
    \lambda_o = - \langle x \rangle + \gamma\frac{\av{d}}{\langle x \rangle},
    \end{equation}
    and
    \begin{equation}
    \lambda_i = -x_i + \frac{ d_i (\langle x \rangle - x_i )}{2(d_i - \av{d})} \pm \frac{d_i \sqrt{(\langle x \rangle - x_i )^2 - 4\gamma^2 x^{-2}(\av{d} - d_i)}}{2(d_i - \av{d})}.
      \label{Regu_zi}
\end{equation}

\begin{figure*}
    \centering
    \includegraphics[width=18cm,height=5cm]{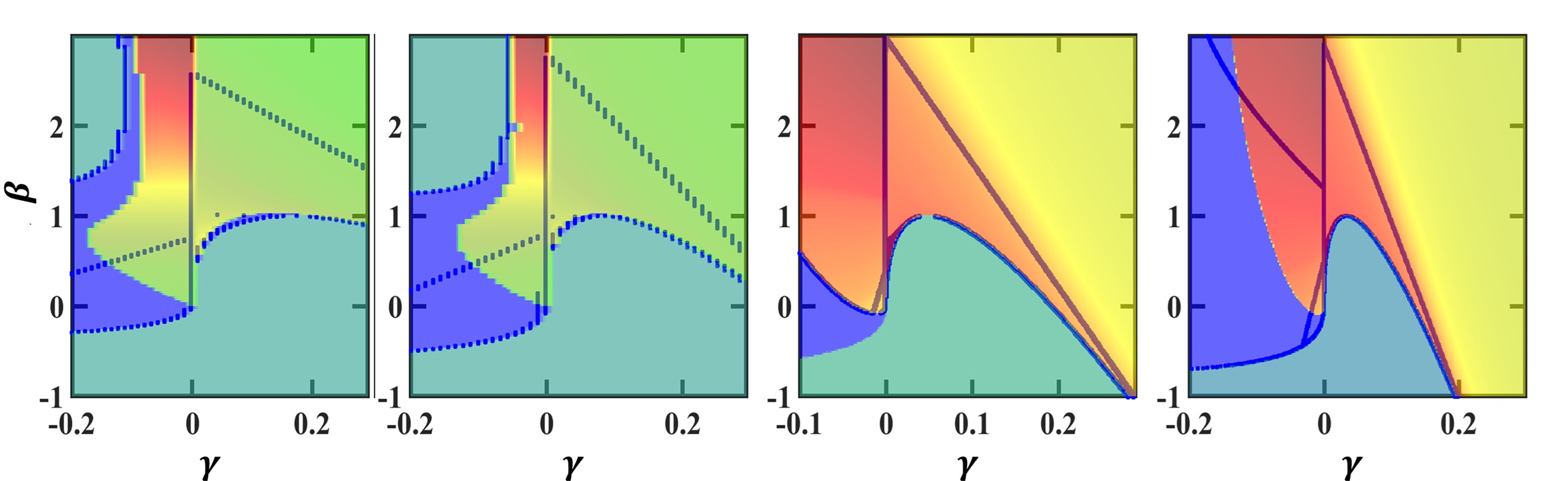}
    \caption{{\bf Regulatory networks. } The first and second figures (from left) correspond to BA networks with average degrees ($\av{d}$) of 6 and 12, respectively (BA1, BA2).    The third and fourth figures represent  (ER) networks (ER1, ER2) with $\av{d} =20$ and $\av{d} =30$, respectively. In the blue and  deep green regime, the system is unstable. The other colors (yellow/yellowish-green to red)  are marked according to the weighted sum of the degree vector and  the  leading eigenvector. In the deep red regime, the large degree ($d_{\rm max}$) plays the key role in determining the largest eigenvalue (calculated from Eqn.\ \ref{Regu_zi}). In the yellow (or yellowish-green) regime, the smallest degree ($d_{\rm min}$) determines the $\lambda_{\max}$ (also calculated from Eqn.\ \ref{Regu_zi}). In the middle, the average degree ($\av{d} $) plays the key role (calculated from Eqn.\ \ref{Regu_zo}). See the  equations\ \eqref{lambda_i_Regu}, and \eqref{lambda_zero_Regu}  in the main text.} 
    \label{fig:Regu_phasespace}
\end{figure*}

\begin{figure*}[t]
    \centering
\includegraphics[width=16cm,height=11cm]{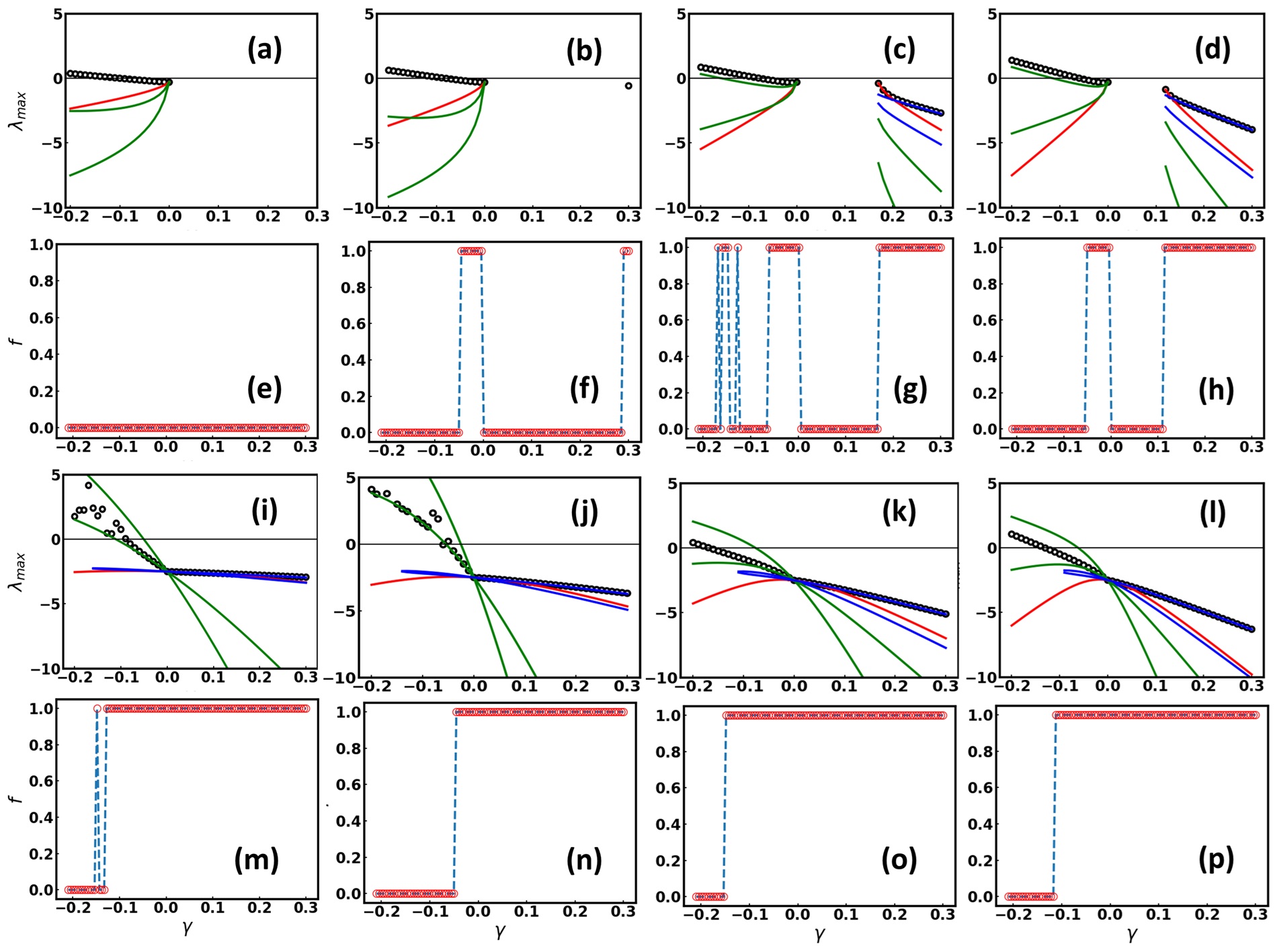}
    \caption{ {\bf Regulatory networks.} We analyze the dynamical nature, and the $\lambda_{\rm max}$ of regulatory networks across different network structures for  varying $\gamma$ values. The first and second columns correspond to BA networks (BA1, and  BA2). The third and fourth columns represent the results of  ER networks (ER1, ER2). In the top two rows, the $\beta$ value is fixed at 0.3, while in the bottom two rows, it is set to 2.5. In subfigures (a)–(d) and (i)–(l), black circles are obtained from numerical simulations.  Green lines indicate the  theoretical lines obtained from 
       Eq.\ \eqref{lambda_i_Regu} (in the main text) mostly captured by the largest degree in the network. The red lines are obtained  from Eq.\ \eqref{lambda_zero_Regu} (in the main text). The  average degrees play the key role here. The blue lines correspond to the minimum degrees  obtained  from Eq.\ \eqref{lambda_i_Regu}  (in the main text). The analytical equations are also written in appendix \ref{regu_net}.
    Subfigures (a)–(d) and (i)–(l) display the $\lambda_{\rm max}$ as a function of $\gamma$. Subfigures (e)–(h) and (m)–(p) illustrate the fraction of nodes with nonzero values, denoted by $f$. These $f$ vs.\  $\gamma$ plots provide supporting evidence for the trends observed in subfigures (a)–(d) and (i)–(l). 
}
    \label{fig:fraction_node}
\end{figure*}

\section{\textbf{Additional figures}}\label{additional figures}
\begin{figure}
    \centering
    \includegraphics[width=0.85\linewidth]{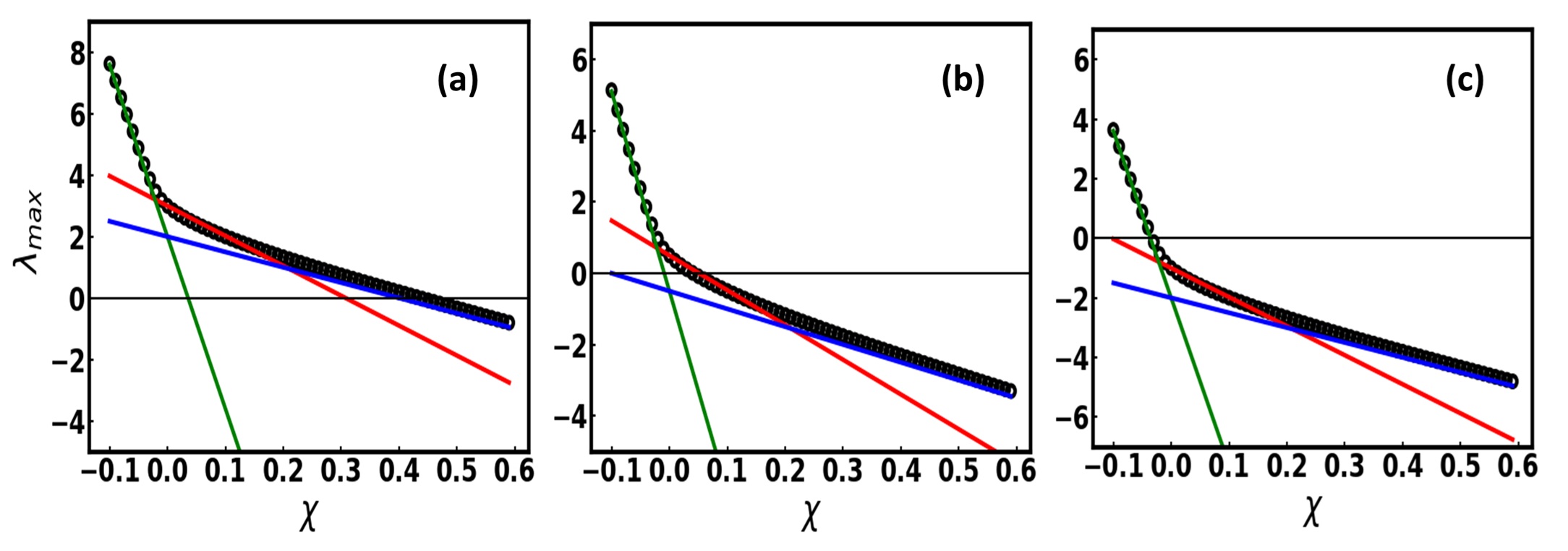}
    \caption{The plot of the $\lambda_{\rm max}$ is shown as a function of $\chi$, varied from -0.1 to 0.6. The parameter $\beta$ is fixed at -2, 0.1, and 2 for Figures (a), (b), and (c), respectively (See figure 1 in the main text). We employed a BA network consisting of 200 nodes with an average degree of 10. The black circles represent values obtained from numerical simulations. Theoretical predictions—depicted by green, red, and blue lines—correspond to the calculations based on the maximum, average, and minimum degrees of the network, respectively, using the  ($\lambda_i$) and ($\lambda_o$)  from the Eq.\ \eqref{lambda_meena} in the main text.   }
    \label{fig:enter-label}
\end{figure}

We have tested our theoretical and numerical results in four different networks. We considered two heterogeneous networks ($N=200$): one is slightly denser compared to the other. In particular, we used Barab{\'a}si-Albert (BA) networks ($P(d)\sim d^{-3}$) with average degrees ($\av{d}$) of 6 and 12. The first BA graph (say BA1)  has a minimum degree of 2 and a maximum degree of 43. The second BA network (say BA2, slightly denser) has minimum degrees of 5 and maximum degrees of 86.  The third and fourth networks (ER1, ER2) are extracted from Poisson distribution  with $\av{d} =20$ and $\av{d} =30$, respectively. For ER1, the minimum degree is set to 10, and the maximum degree is 35. The fourth network (ER2) has a minimum degree of 14 and a maximum degree of 46. 
Figure \ref{fig:SIS_time_sig_inf} presents the stability and mean infection levels across four different networks. The dependence of stability on two key parameters in regulatory networks is illustrated in Fig.\ \ref{fig:Regu_phasespace}. To further probe the system’s behavior, we vary the parameter $\gamma$ and track its effect on both stability and the proportion of nodes that maintain non-zero (feasible) states, as shown in Fig.\ \ref{fig:fraction_node}. Finally, Fig.\ \ref{fig:enter-label} compares the theoretical prediction of the largest eigenvalue, $\lambda_{\rm max}$, with numerical estimates obtained from the dynamic Jacobian ensemble.

\bibliographystyle{apsrev4-1} 
\bibliography{citepaper}

\end{document}